\documentclass[amsmath,amssymb,prl,aps,twocolumn,mathbbm,superscriptaddress]{revtex4-1}
\usepackage{graphicx}
\usepackage{epsfig}
\usepackage[usenames]{color}
\usepackage{epsfig}
\usepackage[normalem]{ulem}
\usepackage{color}
\usepackage[usenames,dvipsnames]{xcolor}
\usepackage[colorlinks=true,citecolor=Cerulean,linkcolor=RubineRed,urlcolor=Cerulean]{hyperref}

\def\la{\langle}
\def\ra{\rangle}

\begin{document}

\title{Dirac Equation in (1+1)-Dimensional Curved Spacetime \\ and the Multiphoton Quantum Rabi Model}

\author{J. S. Pedernales}
\affiliation{Department of Physical Chemistry, University of the Basque Country UPV/EHU, Apartado 644, 48080 Bilbao, Spain}
\affiliation{Institute for Theoretical Physics and IQST, Albert-Einstein-Allee 11, Universit\"at Ulm, D-89069 Ulm, Germany}
\author{M. Beau}
\affiliation{Department of Physics, University of Massachusetts, Boston, MA 02125, USA}
\author{S. M. Pittman}
\affiliation{Department of Physics, Harvard University, Cambridge, MA 02138, USA}
\author{I. L. Egusquiza}
\affiliation{Department of Theoretical Physics and History of Science,\\ University of the Basque Country UPV/EHU, Apartado 644, 48080 Bilbao, Spain}
\author{L. Lamata}
\affiliation{Department of Physical Chemistry, University of the Basque Country UPV/EHU, Apartado 644, 48080 Bilbao, Spain}
\author{E. Solano}
\affiliation{Department of Physical Chemistry, University of the Basque Country UPV/EHU, Apartado 644, 48080 Bilbao, Spain}
\affiliation{IKERBASQUE, Basque Foundation for Science, Maria Diaz de Haro 3, 48013 Bilbao, Spain}
\affiliation{Department of Physics, Shanghai University, 200444 Shanghai, China}
\author{A. del Campo}
\affiliation{Department of Physics, University of Massachusetts, Boston, MA 02125, USA}

\begin{abstract}

We introduce an exact mapping between the Dirac equation in (1+1)-dimensional curved spacetime (DCS) and a multiphoton quantum Rabi model (QRM). A background of a (1+1)-dimensional black hole requires a QRM with one- and two-photon terms that can be implemented in a trapped ion  for the quantum simulation of Dirac particles in curved spacetime. We illustrate our proposal with a numerical analysis of the free fall of a Dirac particle into a (1+1)-dimensional  black hole, and find that the Zitterbewegung effect, measurable via the oscillatory trajectory of the Dirac particle, persists in the presence of gravity. From the duality between the squeezing term in the multiphoton QRM and the metric coupling in the DCS, we show that gravity generates squeezing of the Dirac particle wave function. 

\end{abstract}

\maketitle

{\it Introduction.---} 
The simulation of gravitational theories and related phenomena in the laboratory constitutes an ongoing effort that spans decades of research. Following Unruh's seminal work~\cite{unruh1981experimental} to study Hawking radiation using a sonic analog of a black hole, a variety of systems for the analog simulation of gravity have been put forward. Prominent examples include  classical fluids ~\cite{visser1993acoustic,unruh1995sonic,visser1998acoustic}, shallow water waves~\cite{schutzhold2002, rousseaux2008observation, weinfurtner2011measurement}, Bose-Einstein condensates~\cite{garay2000sonic, barcelo2001analogue, garay2002black, barcelo2003probing, Fisher2003, fedichev2004observer, fischer2004quasiparticle, fischer2004quantum, giovanazzi2004conditions, lahav2010realization, steinhauer2016observation,Fisher2017}, ultracold atoms in optical lattices \cite{rodriguez2017synthetic}, superfluid helium~\cite{volovik1996cosmology,volovik1998induced, jacobson1998event}, nonlinear electrodynamics~\cite{baldovin2000non, novello2000geometrical, de2000light, de2002nonlinear, arellano2006evolving}, slow light~\cite{leonhardt2000relativistic, brevik2001light, leonhardt2002laboratory,unruh2003slow}, waveguides~\cite{schutzhold2005hawking,Angelakis2016}, ion rings~\cite{ horstmann2010hawking}, and laser filaments~\cite{belgiorno2010hawking, unruh2012hawking}; see Refs.~\cite{Barcelo2011,faccio2013analogue} for an extensive review. In the same manner, embedding quantum simulators have recently been identified as suitable candidates for the quantum simulation of Rindler transformations, allowing for the observation of black hole physics and related relativistic phenomena in the lab~\cite{Sabin17}.

In parallel, the field of quantum optics provides a plethora of controllable quantum systems as potential quantum simulators. This has led to a number of analogies between models of quantum optics and other fields of physics. A paradigmatic example is the connection between the quantum Rabi model (QRM), which describes light-matter interaction, and relativistic quantum physics. The simulation of a Dirac fermion in Minkowski spacetime has been proposed and implemented in several platforms~\cite{Lamata07, Gerritsma10, Pedernales15, Weitza2011, Zehnle2017}. However, the connection between quantum optics and quantum theory in curved spacetime remains unexplored.  In this Letter, we complete this missing link by establishing  an analogy between a multiphoton QRM model and a Dirac particle in a (1+1)-dimensional curved spacetime (DCS), which highlights the connection between the two fields. After introducing an exact mapping from a Dirac particle in the background of a (1+1)-dimensional black hole~\cite{Mann91} to the multiphoton QRM, we propose its implementation in a trapped-ion platform. Using numerically exact calculations, we explore the dynamics of a massive Dirac particle in the vicinity of a black hole through the analogy between the multiphoton QRM and DCS. Our results show evidence of the Zitterbewegung effect in the  trajectory of the particle and its density profile. Finally, we demonstrate that gravitation squeezes quantum states as time evolves, in agreement with some recent results~\cite{Jacobson2005,Calzetta03,Schutzhold07,Jacobson2005,Su17}. 

{\it The quantum Rabi model and the Dirac equation.---} The QRM describes the interaction of a two-level atom with a quantized mode of the electromagnetic field. When the wavelength of the  electromagnetic mode greatly exceeds the size of the atom, the dipolar approximation that neglects the spatial dependence of the electromagnetic field justifies  a linear atom-field interaction. Interactions that are quadratic in the field emerge in the description of effective two-level systems due to second-order processes mediated by a virtual third level that is negligibly populated. When the atom-field coupling includes both linear and quadratic terms in the field operators, the Hamiltonian reads $(\hbar=1)$
\begin{equation}\label{QRM1}
H_{\rm R}= \omega  \hat{a}^\dag\hat{a} + \frac{\omega_0}{2}\sigma_z + g \sigma_x(\hat{a} +\hat{a}^\dag) + \kappa \sigma_x (\hat{a}^2 + \hat{a}^{\dag 2}),
\end{equation}
where $\omega$ is the mode frequency, $\omega_0$ the energy splitting of the two level system, and $g$ and $\kappa$ are the coupling strengths of the linear and quadratic terms, respectively. The linear QRM has been proposed and implemented in all its parameter regimes using trapped ions ~\cite{Pedernales15, Dingshun17} and in the ultrastrong and deep-strong coupling regimes using superconducting circuits~\cite{Mezzacapo14, Ustinov16, DiCarlo16}, with protocols that can be as well extended to nonlinear cases~\cite{Felicetti15}. For $\kappa=0$ and $\omega=0$, the corresponding Schr\"odinger equation is equivalent  to the (1+1)-dimensional Dirac equation in flat Minkowski spacetime, $i \partial_t \psi = (mc^2 \sigma_z + p \sigma_x)\psi$, upon identifying $\omega_0/2=mc^2$ and $g=cp_0$. Here $p_0$ is the dimensional part of the momentum operator $p=p_0(a - a^\dag)/i$. This analogy has been exploited in trapped ions for the quantum simulation of relativistic fermions in flat spacetimes~\cite{Gerritsma10, Gerritsma11}. In this Letter, we argue that  the analogy holds when a static gravitational field is included provided that the QRM contains a quadratic two-photon term. 

First, let us recall the general form of the DCS for a fixed metric $g_{\mu\nu}$ in (1+1)-dimensional spacetime. Assume the signature $(+-)$, where $\mu=0$ corresponds to the time component $x^0=c t$ and $\mu=1$ is associated with the space component $x^1=x$. The DCS then reads
\cite{McVittie1932}
\begin{equation}\label{DCSLetter}
\left(i\hbar \gamma^a e_{(a)}^\mu \partial_\mu +\frac{i\hbar}{2}\gamma^a \frac{1}{\sqrt{-g}}\partial_{\mu}\left(\sqrt{-g}\ e_{(a)}^\mu\right)-mc\right)\psi=0\ ,
\end{equation}
where the matrices $\gamma^a$ are given by the standard Pauli matrices $\gamma^0=\sigma_z$  and $\gamma^1=i\sigma_y$, and where $e_{\mu}^{(a)}$ is  a dyad defined as $e_{\mu}^{(a)}=\partial X^{a}/\partial x^{\mu}$, with $X^a$ (resp. $x^\mu$) denoting the $a$ component ($\mu$ component) of the position vector in the Minkowski spacetime (curved spacetime). Dyads satisfy the orthonormality conditions $e_{\mu}^{(a)}e_{(a)}^{\nu}=\delta_{\mu}^{\nu}$.
Now, we consider a semiclassical gravity theory in (1+1) dimensions for a static point source. Notice that in (1+1) dimensions all metrics are conformally flat, and Einstein's equations demand that it be an empty space. There are however interesting modifications of Einstein gravity. In particular we consider the theory in which the curvature is proportional to the trace of the energy momentum tensor~\cite{Mann91, Mann91(2),Mann1992}. The metric is given by $g_{\mu \nu}={\rm diag}[\alpha(x),-1/\alpha(x)]$, with $g_{00}=\alpha(x)=2M|x|+\epsilon$, where $M$ is related to the mass of the point source (in units of inverse length) 
$
M=4\pi G\rho_0 a/c^2\ ,
$
with $\rho_0$ the density and $a$ the spatial distribution radius of the dust, resulting in a total mass for the source of $\rho_0 a$, and where $G$ is the (1+1)-dimensional gravitational constant, which in the Systeme International has units of $\mathrm{kg}^{-1}\mathrm{m}^{1}\mathrm{s}^{-2}$. For the constant  value $\epsilon=+1$, the solution corresponds to the metric induced by a naked source. For $\epsilon=-1$, it corresponds to the exterior black hole solution. 
In this Letter, we show that the DCS in Eq.~\eqref{DCSLetter} for the black hole solution can be exactly mapped onto a multiphoton QRM in Eq. \eqref{QRM1}. Similarly, one can show that the naked source solution can also be mapped to the multiphoton QRM in the weak field approximation $2M|x|\ll 1$, see  \cite{SM}. 

{\it Mapping for the black hole solution.---} Taking $\epsilon=-1$, we have $\alpha(x)=(|x|-r_s)/r_s$, where the corresponding Schwarzschild radius occurs at $r_s=1/(2M)$, which is $r_s = c^2/(4\pi G \rho_0 a)$ in the Systeme International units.  
Since the particle cannot cross the black hole at $x=0$, it is restricted to a region either to the right or to the left of the origin. Here, we restrict the position of the particle to $x>r_s$, which can be done with no loss of generality due to the symmetry of the metric about the origin. This also means that the gravitational redshift factor $\sqrt{g_{00}}=\sqrt{\alpha}$ is restricted to positive values. In order to rewrite Eq. \eqref{DCSLetter}, we introduce operators $\hat{X}$ and $\hat{P}$ to carry out a mapping of the form
 \begin{equation}\label{X}
\hat{X}\equiv r_s\sqrt{\alpha(\hat{x})}\ ,\ 
\hat{P}\equiv -i\hbar \frac{\partial}{\partial X}\ .
\end{equation}
These operators are canonically conjugate and satisfy the commutation relation $[\hat{X},\hat{P}]=i\hbar$.  Under this mapping, the DCS in Eq.~\eqref{DCSLetter} becomes~\cite{SM} 
\begin{equation}\label{BH1}
i\hbar\frac{\partial}{\partial t}\psi = \left(c \sigma_x \frac{1}{4 r_s}\left\{\hat{X},\hat{P}\right\} +  mc^2\sigma_z \frac{\hat{X}}{r_s}\right)\psi\ ,
\end{equation}
where the operator $\{\hat{X},\hat{P}\}$ acts as the generator of squeezing (see below). Alternatively,  Eq.~\eqref{BH1} can be derived choosing the polar coordinates $(X,ct)$, in terms of which the spacetime interval reads $ds^2 = (X^2/r_s^2) c^2 dt^2 - 4dX^2$, see Ref.~\cite{SM}. The new $\hat{X}$ and $\hat{P}$ operators can be mapped to a bosonic field 
\begin{equation}\label{Map2}
\hat{X}=\frac{\lambda}{\sqrt{2}}\left(\hat{a}+\hat{a}^\dagger\right)\ ,\ 
\hat{P}=\frac{\hbar}{i\lambda\sqrt{2}}\left(\hat{a}-\hat{a}^\dagger\right)\ ,
\end{equation}
where $\lambda$  is a constant with units of length. Substituting expressions \eqref{Map2} in Eq. \eqref{BH1} we arrive at $i\hbar\frac{\partial}{\partial t}\psi  =  H_{D}\psi $ with the Hamiltonian 
\begin{equation}\label{BH2}
\hat{H}_{D}=\left(c \sigma_x\frac{1}{4 i r_s }\left(\hat{a}^2- \hat{a}^{\dagger 2}\right)\right.  + \left. mc^2\sigma_z \frac{\lambda}{\sqrt{2}r_s} \left(\hat{a}+\hat{a}^\dagger\right)\right), \ 
\end{equation}
which is formally equivalent to the multiphoton QRM in Eq.~\eqref{QRM1} with  $\omega=\omega_0=0$. Thus, Eq.~\eqref{BH2} encodes the simulation of a Dirac particle in the background of a (1+1)-dimensional black hole.  We point out that the inverse of the Schwarzschild radius $1/r_s$ appears as a multiplicative constant of Hamiltonian \eqref{BH2} and therefore multiplies the time variable in the corresponding unitary evolution operator. As a result, the simulation for a specific value of $r_s$ is tantamount to the simulation for any value of $r_s$ up to  a suitable rescaling of $t$. On the other hand, finding an analogy between the QRM and the DCS in higher dimensions seems a daunting task.

{\it Trapped-ion implementation and numerical tests.---} 
A trapped ion offers suitable quantum degrees of freedom for the simulation of Eq.~(\ref{BH2}), with its mechanical modes behaving as quantum harmonic oscillators that can hold the Hilbert space associated with operators $a$ and $a^\dag$, and two of its electronic states implementing the Hilbert space associated with Pauli operators.

To simplify the implementation, we change the $\sigma_z$ Pauli operator in the second term of Eq. \eqref{BH2} into a $\sigma_y$, without altering the physics of the model.  We propose to implement the term $ [mc^2\lambda/(\sqrt{2} r_s)] \sigma_y(a+a^\dagger)\psi$ with red and blue sideband interactions, using Hamiltonians $-i\eta\Omega_r (\sigma^+ a-\sigma^- a^\dag)$ and $i\eta\Omega_b (\sigma^+ a^\dag - \sigma^- a)$, respectively.  
The corresponding Rabi frequencies are
\begin{equation}
\eta \Omega_r  =   mc^2\lambda/(\sqrt{2} r_s)\ ,\ 
\eta \Omega_b  =  - mc^2\lambda/(\sqrt{2}r_s)\ .
\end{equation}
Similarly, the term $[\hbar c/(4ir_s)]\sigma_x\left(a^2-(a^\dagger)^2\right)\psi $ can be implemented with red and blue second sidebands, $-i\eta_2^2\Omega_{r,2} (\sigma^+ a^2-\sigma^- (a^\dag)^2)$ and $i\eta_2^2\Omega_{b,2} (\sigma^+ (a^\dag)^2 - \sigma^- a^2)$, with $\eta_2^2 \Omega_{r,2}=\eta_2^2 \Omega_{b,2}= \hbar c/(4r_s)$. Note that the values of Rabi frequencies $\Omega_{\rm r(b)}$ and Lamb-Dicke parameters $\eta$ for first and second sidebands can be set individually, given that these will be excited with independent laser fields, and therefore the ratio between first and second sideband interaction strengths can be set at will.

The position of the simulated Dirac particle at time $t$ can be associated with observables of the mechanical degrees of freedom of the ion through the equivalence
$
\frac{1}{r_s}\hat{X}(t)^2\equiv \hat{x}(t) - r_s\ ,
$
where $\hat{x}(t)$ and $\hat{X}(t)^2$ are given in the Heisenberg representation. Similarly, the position of the ion moving in the trap can be associated with the redshift factor multiplied by $r_s$ and with the position of the Dirac particle using the polar coordinates $(t,X)$ as mentioned above. From an experimental point of view, the position of a trapped ion as well as higher order moments can be measured by mapping the information of the motional state of the ion to its internal degrees of freedom. Such measurements suffice to reconstruct the density profile of the ion as done in Refs.\cite{Zahringer10, Gerritsma10, Gerritsma11}.   

\begin{figure}[t]
\begin{center}
\includegraphics[width=0.9\columnwidth]{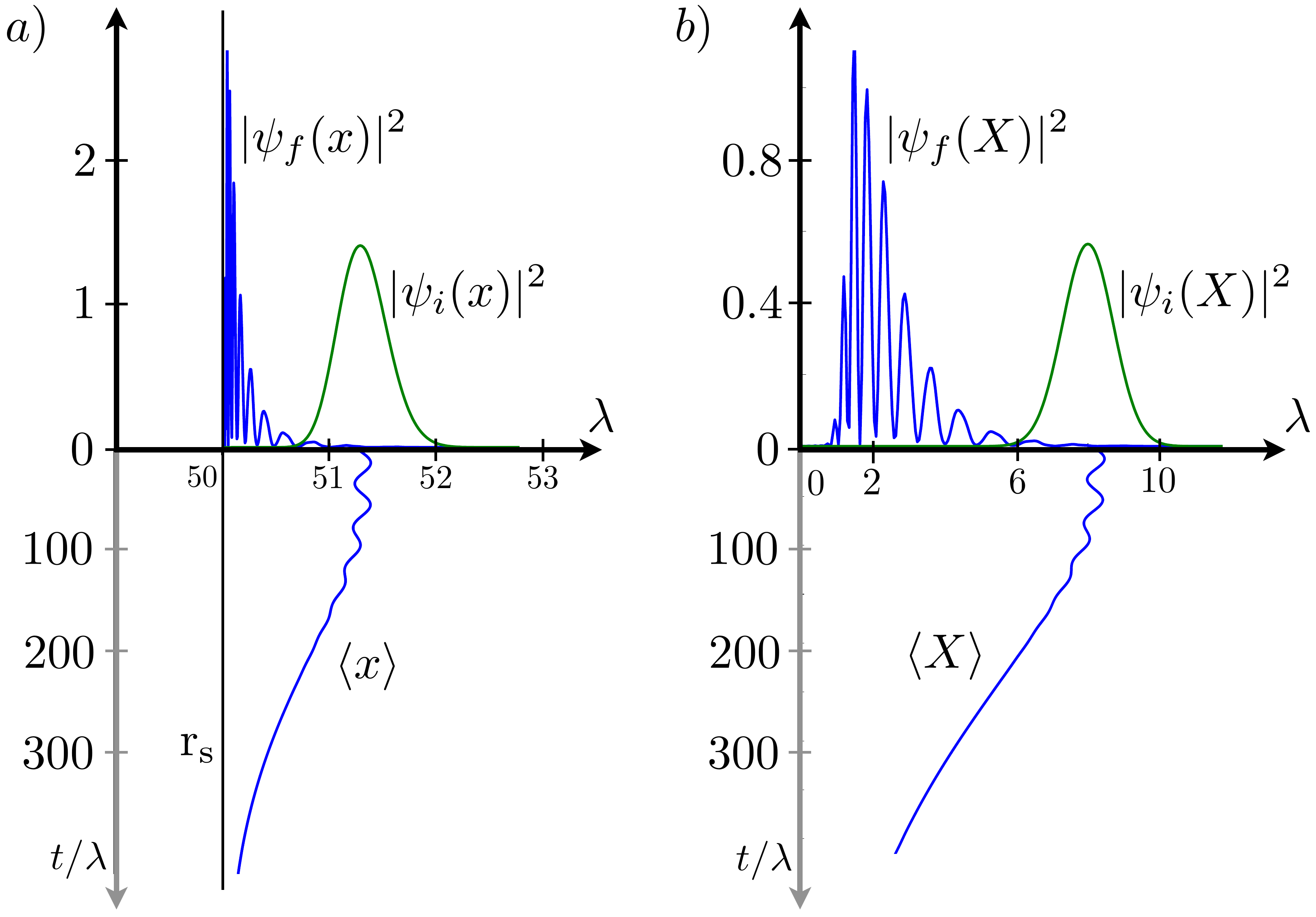}
\caption{{\bf Dynamics of a massive Dirac particle near a black hole  and the multiphoton QRM.} 
Properties of the simulated Dirac particle (a) are related to those of the ion (b), under the mapping~(\ref{X}). 
In both cases the upper plot shows the initial (green) and final (blue) probability density profiles, while the lower plots show the expectation values of the corresponding position operators. For the simulated Dirac particle, the position of the horizon is indicated by a vertical line labeled $r_s$. The initial state of the ion is $\Psi_0=\phi_{X_0/\lambda=8} (X)\otimes | + \rangle_x$, where $\phi_{X_0/\lambda=8}(X)$ is a Gaussian wave function centered at $X_0$, and $| + \rangle_x$ is the eigenstate of operator $\sigma_x$ with positive eigenvalue. The simulation is performed under Hamiltonian in Eq.~\eqref{BH2}, for the case where $m\lambda=0.3$ and $M\lambda=0.01$, with $c=1$. The scale of the oscillations of the simulated particle is smaller than that of the oscillations of the ion, as the position of the simulated particle is rescaled by $1/r_{\rm s}$ under mapping in Eq.~\ref{X}.}
\label{massive}
\end{center}
\end{figure}

We consider the initial state $|\Psi_0\ra=|\phi_0\ra \otimes |{\chi}\ra$, where $|\phi_0\ra$ and $|{\chi}\ra$ are, respectively, the wave functions of the spatial and internal degrees of freedom of the ion, which are unentangled at time $t=0$. We numerically simulate the unitary evolution of such a state under Hamiltonian~(\ref{BH2}) and track the expectation value  $\langle X^2 \rangle (t)$, related to the mechanical degrees of freedom of the ion. From it, we compute the expectation value of the position operator of the simulated Dirac particle, $x$. Note that for the semiclassical approximation to hold, the Compton wavelength of the Dirac particle $\lambda_c=h/(mc)$ must be much smaller than the Schwarzschild radius, $\lambda_c\ll r_s$.
In Fig.~\ref{massive} we show numerical results for the case of a massive particle; specifically, we analyze the regime in which $m = 0.3/\lambda$ and $M=0.01/\lambda$, with $c=1$. The ion is initialized with its internal state in $| + \rangle_x $, where $|\pm\rangle_x$ are eigenstates of $\hat{\sigma}_x$. A Gaussian distribution $\phi_{X_0}(X)\equiv \langle X|\phi_{0}\rangle=\mathcal{N}e^{-\frac{(X-X_0)^2}{2\sigma^2}}$ describes the initial state of the mechanical degrees of freedom localized in half-space $X>0$, with values $X_0/\lambda=8$ and $\sigma/\lambda=1$ corresponding to a vacuum state displaced by $\alpha / \lambda =8$, and where $\mathcal{N}$ denotes the normalization factor.  The trapped ion can be prepared in such an initial motional state by simultaneously applying resonant red and blue sidebands on it, when it is in its lower energy state~\cite{Meekhof96}, or via Bang-Bang techniques~\cite{Alonso16}. After changing the space variable $X(x)\equiv r_s\sqrt{\alpha(x)}=r_s\sqrt{(x/r_s)-1}$, we find that the corresponding initial wave function localized in the region $x>r_s$ for the simulated Dirac particle is given by $\phi_{x_0}(x)\equiv \langle x|\phi_{0}\rangle=[\partial X(x)/\partial x]^{1/2}\phi_{X_0}(X(x))$, where $x_0\equiv (X_0^2/r_s)+r_s$.    

{\it Zitterbewegung effect in the presence of gravity.---}
As shown in Fig. \ref{massive}(a), the simulated Dirac particle approaches asymptotically the horizon of the black hole at $x=r_s$. As the validity of our mapping is preserved throughout the entire time evolution,  the ion does not cross the origin $X=0$, see Fig. \ref{massive}(b). In addition, the ion trajectory exhibits an oscillatory behavior,  with an amplitude that vanishes when the particle approaches the horizon. We associate this phenomenon with the Zitterbewegung effect, well known for massive relativistic fermions in flat spacetime, and originating from the interference between positive and negative energy solutions of the Dirac equation. 
We show that such a phenomenon persists in the presence of gravity. Intuitively, one can argue that this is indeed the case as a curved spacetime can be described locally by a Minkowski metric, in which we know that the particle manifests the Zitterbewegung effect. The equation of motion for the expectation value of the position operator $\hat{x}(t)$ reads \cite{SM}
\begin{eqnarray}\label{SM:SecondDer_x(t)}
\frac{d^2}{dt^2}\langle \hat{x}(t) \rangle&=& \frac{c^2}{r_s}\left(\frac{\langle \hat{x}(t)\rangle}{r_s}-1\right) 
\\ & & 
- \frac{2mc^3 }{\hbar}\left\langle \Big(\frac{\hat{x}(t)}{r_s}-1\Big)^{3/2} e^{2 i \hat{H}_{D} t/\hbar}  \hat{\sigma}_y\right\rangle\ ,\nonumber
\end{eqnarray}
where the second term, which depends on the mass, induces the oscillations in Fig. \ref{massive}(a).
In Ref.~\cite{SM} we show that the amplitude of the oscillations decreases as the particle approaches the horizon $\langle \hat{x}(t)\rangle\rightarrow r_s$, see also Fig. \eqref{massive}(a). 
Our mapping offers an alternative way to observe the Zitterbewegung effect based on the recorded values of the redshift factor $\langle\hat{X}(t)\rangle/r_s$, see Fig. \ref{massive}(b). Indeed, oscillations between red and blue shifts  provide a direct signature of the Zitterbewegung effect. Note however that for a massless particle $m=0$ this term vanishes, suppressing the Zitterbewegung effect, as expected ~\cite{Lamata07,SM}. In the massless case, an ion initialized with internal state $| +(-)\rangle_x$, i.e., in the positive (negative) chirality, moves away from the origin (towards the horizon). 

Figure~\ref{massive} further shows an interference pattern in the density profile that appears all along the dynamics. We identify  this phenomenon as an additional signature of the Zitterbewegung effect. Indeed, this can be understood as an interference between positive and negative energy solutions of the Dirac equation that persists at {\it long time}. 
In flat spacetime positive and negative energy solutions spread in opposite directions and therefore do not overlap at long times. However, in the presence of gravity the two solutions approach the horizon (without crossing it). 
This results in the spatial squeezing of the density profile of the particle shown in Fig. \ref{massive}. The overlap between both positive and negative energy solutions is therefore maximized as the particle approaches the horizon and the oscillations in the trajectory are suppressed. 
We note that for the massless case the interference pattern in the density profile is absent (see Ref.~\cite{SM}), consistently with the suppression of the Zitterbewegung effect.

\begin{figure}[t]
\begin{center}
\includegraphics[width=0.8\columnwidth]{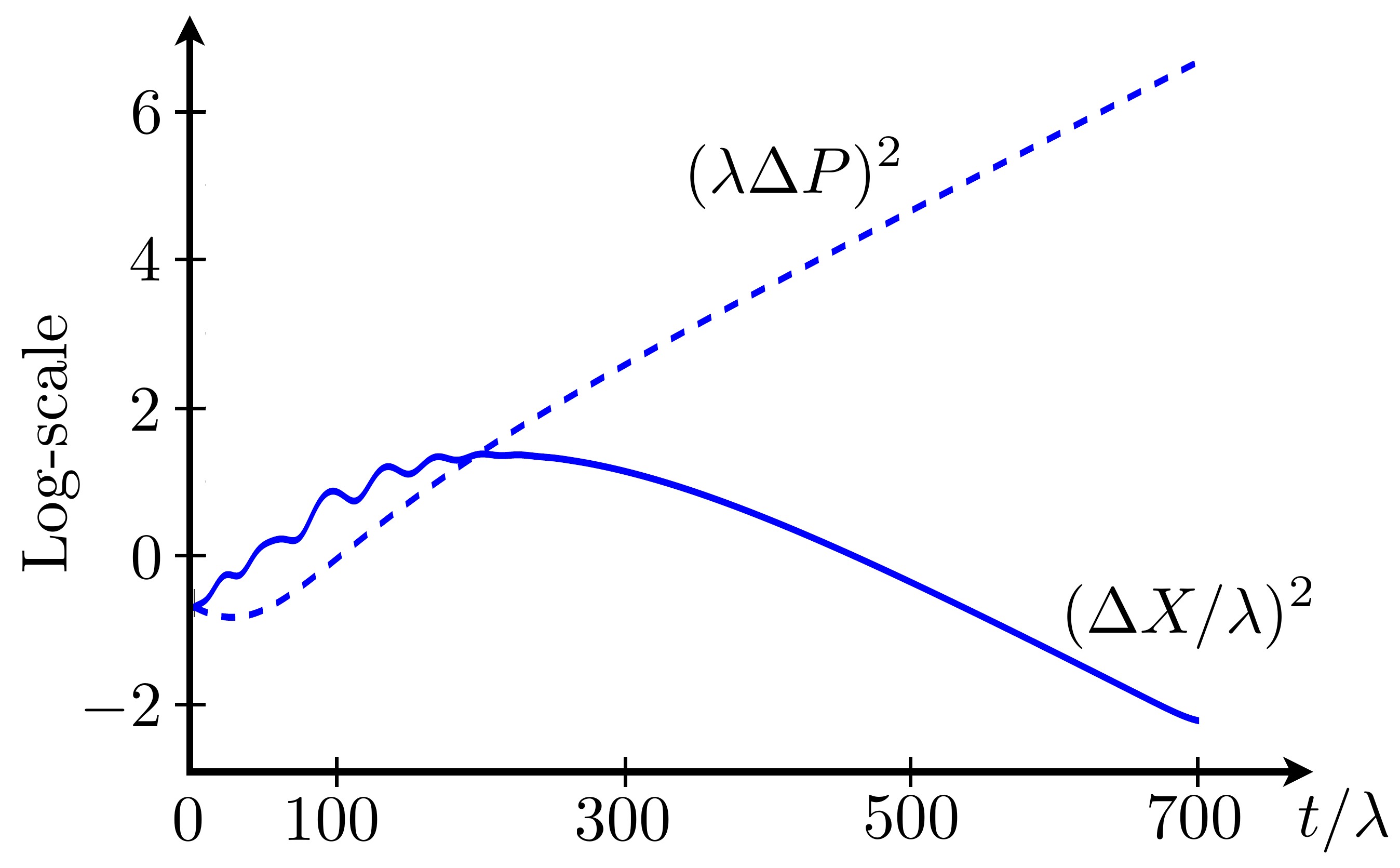}
\caption{{\bf Squeezing of the dynamics by curvature of spacetime. }  The continuous line corresponds to the variance of the position of the ion $(\Delta X/\lambda)^2$ while the dashed line corresponds to the variance of the momentum of the ion $(\lambda \Delta P)^2$, both  being dimensionless. A logarithmic scale is used. The simulation regime is the same as that in Fig.~\ref{massive}. As the ion evolves in time a clear trend towards its position getting localized accompanied with an exponential growth of the uncertainty of its momentum can be observed.} 
\label{Squeezing}
\end{center}
\end{figure}
{\it Squeezed states and gravity.---} To understand the spatial squeezing of the density profile (see Fig. \ref{massive}), we stress that the mapping we introduce, see Eqs. \eqref{BH1}-\eqref{BH2}, suggests an analogy between \textit{squeezing} in quantum optics and \textit{curvature} of spacetime in the context of the relativistic Dirac equation.  From this analogy, we expect that gravity generates squeezed states as time evolves. A standard way to characterize squeezing along the dynamics is to look at the time evolution of the variance of the position and the momentum, $\Delta X(t)$ and $\Delta P(t)$ [defined for an arbitrary operator $O$ as $\Delta O(t) \equiv \sqrt{\la \hat{O}(t)^2 \ra-\la \hat{O}(t)\ra^2}$].
In Fig.~\ref{Squeezing} we show that the variance of the momentum of the ion $\Delta \hat{P}(t)$ grows indefinitely when the particle approaches the horizon of the black hole while the variance of the position $\Delta \hat{X}(t)$ decreases and tends asymptotically to zero. This gives a signature of the squeezing of the wave function when the particle approaches the horizon. A similar squeezing is observed as well in the massless case, as we show in Ref.~\cite{SM}.  
Note that the relation between squeezing and gravity has been recently explored in the context of cosmological particle creation \cite{Jacobson2005}, in analogues using Bose-Einstein condensates \cite{Calzetta03, barcelo2003probing, Fisher2004} and trapped ions \cite{Schutzhold07}, and in relation to the Hawking effect near a Schwarzschild black hole \cite{Jacobson2005,Su17}. However, our results exploit the first-quantisation formalism, and therefore the predicted squeezing relates exclusively to the phase space of the simulated particle and cannot be, a priori, associated with cosmological particle creation.

{\it Conclusions.---} We have proposed an analogy between the DCS and a multiphoton QRM. We have shown that the former can be exactly mapped to the latter when the metric describes a (1+1)-dimensional analogue of a Schwarzschild black hole. We have proposed the implementation of the mapping with a single trapped ion and used numerical results to illustrate the dynamics of the ion and a simulated Dirac particle. Our results show that the Zitterbewegung effect in curved spacetime leads to the oscillatory trajectory of the ion and  the interference pattern in its probability density profile. In addition, our findings demonstrate that gravitation and quantum squeezing are strongly related and we hope that our present work will motivate further research in this direction. The analogy presented here illustrates a connection between relativistic quantum mechanics in curved spacetimes and basic light-matter interaction models, which may inspire quantum simulations of relativistic equations in curved spacetimes in a variety of quantum platforms.

We acknowledge support from Spanish Ministerio de Econom\'ia y Competitividad/Fondo Europeo de Desarrollo Regional FIS2015-69983-P, Basque Government IT986-16, Ram\'on y Cajal Grant No. RYC-2012-11391, UMass Boston (Project No. P20150000029279), and the John Templeton Foundation.

\bibliography{DCSBib2} 
\bibliographystyle{apsrev4-1}

\cleardoublepage
\onecolumngrid

{\begin{center} \large SUPPLEMENTAL MATERIAL \end{center}}

\section{Dirac equation in Curved Spacetime (DCS) and quantum Rabi model (QRM)}\label{SM:DCS-QRM}

\subsection{Dirac equation in Curved Spacetime (DCS)}\label{SM:DCS-QRM:DCS}

We consider a fixed metric $g_{\mu\nu}$ in a (1+1)-dimensional spacetime with the signature $(+-)$, where $\mu=0$ corresponds to the time component $x^0=c t$ and $\mu=1$ is associated with the space component $x^1=x$. The Minkowski metric $\eta_{ab}$ is then given by 
\begin{equation}
\eta_{ab}=\begin{pmatrix}
1 & 0\\
0 & -1
\end{pmatrix}\ ,
\end{equation}
where we associate the Latin indices ($a,b,\dots$) with the Minkowski metric and the Greek indices ($\mu,\nu,\dots$)  with the curved metric. 
The relation between Minkowski and curved metric is 
\begin{equation}
g_{\mu\nu} = e_{\mu}^{(a)}e_{\nu}^{(b)}\eta_{ab}\ ,
\end{equation}
where $e_{\mu}^{(a)}$ is  a dyad defined as $e_{\mu}^{(a)}=\partial X^{a}/\partial x^{\mu}$ with $X^a$ ($x^\mu$) denoting the $a$ component ($\mu$ component) of the position vector in the Minkowski spacetime (curved spacetime). Dyads satisfy the orthonormality condition 
$$e_{\mu}^{(a)}e_{(a)}^{\nu}=\delta_{\mu}^{\nu}\ .$$

It is known that the DCS is given by
\begin{equation}\label{DCS1}
\left(i\hbar\gamma^\mu \nabla_\mu + mc\right) \psi =0\ ,
\end{equation}
where the $\gamma$ matrices satisfy the relations (Clifford algebra)
\begin{align}
&\left\{\gamma^{a},\gamma^{b} \right\}=2\eta^{ab}, \\
&\left\{\gamma^\mu,\gamma^\mu \right\}=2g^{\mu\nu}\ ,
\end{align}
leading to
\begin{equation}
\gamma^\mu = e_{(a)}^{\mu}\gamma^{a}\ ,
\end{equation}
with the matrices $\gamma^a$ being defined as
\begin{align}\label{gamma}
 \gamma^0=\sigma_z=\begin{pmatrix}
1 & 0\\
0 & -1
\end{pmatrix}\ ,\ 
 \gamma^1=i\sigma_y=\begin{pmatrix}
0 & 1\\
-1 & 0
\end{pmatrix}\ .
\end{align}
In \eqref{DCS1} the covariant derivative $\nabla_\mu$ reads
\begin{equation}
\nabla_\mu=\partial_\mu + \Gamma_\mu,
\end{equation}
where $\Gamma_\mu$ is the spinor connection equal to $\frac{1}{2}\sigma^{bc}\omega_{bc\mu}$. In the latter expression, we give the generator of Lorentz rotations $\sigma^{bc}=\frac{1}{4}[\gamma^a,\gamma^b]$, with $\omega_{bc\mu}=e_{(b)}^\nu D_\mu e_{(c)\nu}$, where the standard covariant derivative is given by $D_\mu A_\nu = \partial_\mu A_\mu +\Gamma^{\sigma}_{\mu\nu}A_\sigma$ for any covariant vector $A_\mu$. Here, we introduce the Christoffel symbol $\Gamma^{\sigma}_{\mu\nu} = \frac{1}{2}g^{\sigma\rho}\left(\partial_{\mu}g_{\nu\rho}+\partial_{\nu}g_{\mu\rho}-\partial_\rho g_{\mu\nu}\right)$.
Gathering the previous results, the following explicit form of the DCS is obtained (see \cite{Mann91})
\begin{equation}\label{DCS2}
\left(i\hbar \gamma^a e_{(a)}^\mu \partial_\mu +\frac{i\hbar}{2}\gamma^a \frac{1}{\sqrt{-g}}\partial_{\mu}\left(\sqrt{-g}\ e_{(a)}^\mu\right)-mc\right)\psi=0\ ,
\end{equation}
where the scalar $g$ is the determinant of the metric $g_{\mu\nu}$. 
Equation \eqref{DCS2} is the (1+1)-dimensional DCS. For higher dimension, the spinor connection cannot be simplified as easily and one finds a more complicated expression \cite{McVittie1932}.

\subsection{Diagonal metric}\label{SM:DCS-QRM:diagmetric}

In what follows we consider a diagonal metric given by \cite{Mann91}
\begin{equation}\label{metricg}
g_{\mu\nu}=\begin{pmatrix}
\alpha(x) & 0\\
0 & -1/\alpha(x)
\end{pmatrix}\ ,
\end{equation}
where $\alpha(x)$ is a non-zero function of the position $x$. Notice that the Jacobian $\sqrt{-g(x)}=1$, where $g(x)$ is the determinant of the metric.  
It is clear that for the metric in \eqref{metricg}, we have
\begin{align}\label{Dyads}
& e_{(0)}^{0}=\frac{1}{\sqrt{\alpha(x)}}\ ,\ e_{(1)}^{1}=\sqrt{\alpha(x)}\ ,\ e_{(1)}^{0}=e_{(0)}^{1}=0\ .
\end{align}
This equation gives the elements $e_{(a)}^\mu$, where the Latin indices appear in the bottom as in Eq.~\eqref{DCS2}.

In Eq.~\eqref{DCS2}, we arbitrarily chose $\gamma^0=\sigma_z$ and $\gamma^1=i\sigma_y$, see Eq.~\eqref{gamma}, and $mc\mathbb{I}$ for the rest energy term of the Hamiltonian. However, we could have chosen a different set of $\gamma$-functions satisfying the Clifford algebra by taking $\gamma^0=\mathbb{I}$, $\gamma^1=-\sigma_x$, and $mc\sigma_z$ instead. It is clear that the mapping between these two representations is given by a multiplication of \eqref{DCS2} by the Pauli matrix $\sigma_z$ (we use $\sigma_z\sigma_y=-i\sigma_x$ and multiply by $+i$). Therefore, for a diagonal metric \eqref{metricg} we obtain
\begin{equation}\label{DCS3}
\frac{1}{\sqrt{\alpha(x)}}\ i\hbar\frac{\partial}{\partial t}\psi = \left( c \sigma_x \sqrt{\alpha(x)}\  (-i\hbar)\frac{\partial}{\partial x} +\frac{c\sigma_x}{2}\ (-i \hbar)\frac{\partial}{\partial x}\left(\sqrt{\alpha(x)}\right)+mc^2\sigma_z\right)\psi\ .
\end{equation}
 
Let us rewrite the following operator 
\begin{equation}\label{CovD1}
\hat{D}\equiv\sqrt{\alpha(x)}\  (-i\hbar)\frac{\partial}{\partial x} +\frac{1}{2}(-i \hbar)\frac{\partial}{\partial x}\left(\sqrt{\alpha(x)}\right)\ .
\end{equation}
Now, we define $\hat{f}= f(\hat{x})\equiv\sqrt{\alpha(\hat{x})}$ and recall the expression for the momentum operator $\hat{p}\equiv -i\hbar\partial/\partial x$. It is clear that $\hat{f}\hat{p}$ is not Hermitian. However,  the operator $\hat{A}\equiv\hat{f}\hat{p}-\frac{1}{2}[\hat{f},\hat{p}]=\frac{1}{2}\left\{\hat{f},\hat{p}\right\}$ is a Hermitian operator by construction. Given that $\hat{f}$ is a function of the position operator $\hat{x}$, we note that $[\hat{f},\hat{p}]\psi=i\hbar f'(x)\psi$, where $f'(x)\equiv \partial f(x)/\partial x$. Hence, Eq. \eqref{CovD1} can be rewritten as
\begin{equation}\label{CovD3}
\hat{D}=\frac{1}{2}\left\{\sqrt{\alpha(\hat{x})},\hat{p}\right\}\ .
\end{equation}   
Finally, we find that the DCS reads
\begin{equation}\label{DCS4}
\frac{1}{\sqrt{\alpha(\hat{x})}}\ i\hbar\frac{\partial}{\partial t}\psi = \left(c \sigma_x\frac{1}{2}\left\{\sqrt{\alpha(\hat{x})},\hat{p}\right\} + mc^2\sigma_z\right)\psi\ .
\end{equation}

\subsection{(1+1)-dimensional black hole}\label{SM:DCS-QRM:BH}

As discussed in the main text,  in the theory of (1+1)-dimensional gravity we consider, the field equations for a point source admit a black hole solution~\cite{Mann91, Mann91(2),Mann1992} for which the diagonal metric is given by Eq.~\eqref{metricg} with 
\begin{equation}
\alpha(x) = 2M|x| - 1 = \frac{|x|}{r_s}-1 \ , 
\label{alpha}
\end{equation}
where $r_s=1/(2M)$ is the Schwarzschild radius.  

Now we introduce the operators $\hat{X}$ and $\hat{P}$ 
 \begin{subequations}\label{SM:XP}
\begin{equation}\label{SM:X}
\hat{X}\equiv r_s\sqrt{\alpha(\hat{x})}\ ,
\end{equation}
\begin{equation}\label{SM:P}
\hat{P}\equiv -i\hbar \frac{\partial}{\partial X}\ ,
\end{equation}
\end{subequations}
where $X=r_s\sqrt{\alpha(x)}$ is an eigenvalue of the operator $\hat{X}$, and rewrite the anticommutator in Eq.~\eqref{DCS4} as
\begin{equation}
\left\{\sqrt{\alpha(\hat{x})},\hat{p}\right\}  =  2\sqrt{\alpha(x)} \hat{p} -i\hbar \frac{\partial}{\partial x}(\sqrt{\alpha(x)}) = -2i\hbar\sqrt{\alpha(x)}\frac{\partial (\sqrt{\alpha(x)})}{\partial x}\frac{\partial}{\partial\sqrt{\alpha}}-i\hbar \frac{\partial}{\partial x}(\sqrt{\alpha(x)})\ ,
\end{equation}
with $\partial\sqrt{\alpha}/\partial x = M/\sqrt{\alpha}=1/(2r_s\sqrt{\alpha})$, leading to
\begin{equation*}
\left\{\sqrt{\alpha(\hat{x})},\hat{p}\right\} = -i\hbar\frac{1}{r_s} \frac{\partial}{\partial\sqrt{\alpha}}-i\hbar \frac{1}{2r_s\sqrt{\alpha(\hat{x})}}= \hat{P}-i\hbar \frac{1}{2\hat{X}}\ .
\end{equation*}
The operator $1/(2M\hat{X})=1/\sqrt{\alpha(\hat{x})}$ appears in left hand side of Eq.~\eqref{DCS4}. After multiplying both sides by $2M\hat{X}$, we obtain 
\begin{equation}\label{SM:BH1}
i\hbar\frac{\partial}{\partial t}\psi = \left(c \sigma_x \frac{1}{4r_s}\left\{\hat{X},\hat{P}\right\} + mc^2\sigma_z \frac{\hat{X}}{r_s}\right)\psi,
\end{equation}
where we have used the identity $2\hat{X}\hat{P}-i\hbar  = \left\{\hat{X},\hat{P}\right\}$.

\textit{Second derivation.} 

An alternative derivation of  Eq.~(\ref{SM:BH1}) exploit the  choice of the polar coordinates $(X,ct)$, in terms of which  the spacetime interval is given by $ds^2 = (X^2/r_s^2) c^2 dt^2 - 4dX^2$. It follows that
\begin{equation*}
g_{\mu\nu}(X)=\begin{pmatrix}
X^2/r_s^2 & 0\\
0 & -4
\end{pmatrix}\ .
\end{equation*}
The Jacobian equals $\sqrt{-g}=2|X|/r_s$ and the dyads are given by
\begin{align}\label{Dyads}
& e_{(0)}^{0}=\frac{r_s}{|X|}\ ,\ e_{(1)}^{1}=\frac{1}{2}\ ,\ e_{(1)}^{0}=e_{(0)}^{1}=0\ .
\end{align}
Using these identities in Eq. \eqref{DCS2} and multiplying both sides of the equation by $X/r_s$ we obtain Eq. \eqref{SM:BH1}.

\textit{Time-dependent metric.}

One could also consider an $\alpha(x,t)$ that not only does depend on $x$, but also on time, that is to say a dynamic metric. In this case Eq.~(\ref{DCS2}) transforms into
\begin{equation}
i\hbar \frac{1}{\sqrt{\alpha(x,t)}} \partial_t \psi = [ -i\hbar \{c \sigma_x \sqrt{\alpha(x,t)}\partial_x + \frac{1}{2}c\sigma_x \partial_x (\sqrt{\alpha(x,t)}) + \frac{1}{2}\partial_t \left( \frac{1}{\sqrt{\alpha(x,t)}} \right) \} + mc^2\sigma_z ] \psi,
\end{equation}
which contains one additional term with respect to Eq.~(\ref{DCS3}), namely the third term in the right-hand side of the equation. Without an a priori justification from a theory of gravity, the experimental implementation of the DCS with a trapped ion readily accommodates the possibility of considering a time-dependent metric of the form $\alpha(x,t)=\alpha(x)\beta(t)$, with the same $\alpha(x)$ as in the previous section (Eq.~(\ref{alpha})) and the same mapping as in Eq.~(\ref{SM:XP}). The corresponding DCS is given by
\begin{equation}
i\hbar\frac{\partial}{\partial t}\psi = \left(\beta(t) c \sigma_x \frac{1}{4r_s}\left\{\hat{X},\hat{P}\right\}  + \sqrt{\beta(t)} mc^2\sigma_z \frac{\hat{X}}{r_s} - i\hbar \frac{\beta'(t)}{2 \beta(t)}\right)\psi.
\end{equation}
The third term in the right-hand side of the Schr\"odinger equation can be ignored, as it will result in an unobservable global phase. Regarding trapped ion implementation, the time dependence of the first and second terms can always be accounted for absorbing it in the Rabi frequencies of the first and second sidebands, which can be made time dependent by controlling the laser intensity.

\section{Naked source: weak gravity field - constant acceleration field}\label{SM:NakedSource}

In this section, we consider that the gravity field is induced by a ``naked source'', meaning that there is no black hole solution. These two models are physically relevant in 3D if one considers that $g\equiv -M c^2<0$ is a constant acceleration field obtained for a weak gravity field in a small region of the space, that is, for $M|x|\ll 1$. The corresponding classical Newtonian potential is given by $V(x)=-mg|x|<0$ confining the particle. For a particle moving in a small region of space with $x>0$ we have $V(x)=-mgx<0$ modeling the particle free fall.   

Here, we consider a small perturbation of the Minkowski metric, with  $M\ll 1$ (i.e., $\sqrt{\alpha(x)}\approx 1+M |x|$). We also assume that the particle moves in a small region of space $M|x|\ll 1$ such that we can take $x>0$. After multiplying both sides of Eq.~(\ref{DCS2}) by $\sqrt{\alpha}$ and taking the first order expansion in $M$, we find
\begin{eqnarray}
i\hbar\frac{\partial}{\partial t}\psi & = &  c \sigma_x\left(\hat{p}+\frac{M}{2}\left\{\hat{x},\hat{p}\right\}\right)\psi
+ mc^2\sigma_z(1+M \hat{x})\psi\ .\label{DCS5}
\end{eqnarray}

Let us now introduce the standard creation and annihilation operators that satisfy $[a,a^\dagger]=1,\ [a,a]=[a^\dag,a^\dag]=0$. In terms of them, 
\begin{eqnarray}\label{SM:Map1}
\hat{x}  =  \frac{\lambda}{\sqrt{2}}\left(a+a^\dagger\right)\ , \,\, \hat{p} =  \frac{\hbar}{i\lambda\sqrt{2}}\left(a-a^\dagger\right)\ ,
\end{eqnarray}
where the length scale $\lambda=\sqrt{\hbar/(m\nu_x)}$ is related to the frequency of the ion trap $\nu_x$ and $m$ is the mass of the ion used for the quantum simulation of the DCS. In this representation, 
\begin{eqnarray}
\left\{\hat{x},\hat{p}\right\}=\frac{1}{2i}\left(a^2-(a^\dagger)^2\right)\ .
\end{eqnarray}
Therefore, we can find the map between DCS and a modified QRM,
\begin{equation}\label{SM:DCS9}
i\hbar\frac{\partial}{\partial t}\psi =  \left(c \sigma_x\frac{i\hbar}{\lambda\sqrt{2}}\left(a^\dagger-a\right)+\hbar c \sigma_x\frac{M}{4i}\left(a^2-(a^\dagger)^2\right)\right)\psi + mc^2\sigma_z\left(1+\frac{\lambda M}{\sqrt{2}}\left(a+a^\dagger\right) \right)\psi\ ,
\end{equation}
corresponding to a one- and two-photon QRM with an additional term equal to the product between the mass and the position operator multiplied by a constant. Surprisingly, Eq. \eqref{SM:DCS9} is equivalent to Eq.~(6) in the main text, up to a displacement of the wave packet $X\mapsto X-X_d$, where $X_d=-1/(2M)$. To show this, it suffices to displace the solution of Eq.~(6) in the main text using the operator $D_{\alpha}=\exp{\left(\alpha \hat{a}-\alpha^\ast\hat{a}^\dagger\right)}$, where $\alpha = \sqrt{2} X_d/\lambda$. Therefore, the two mappings, that of the black hole and the one of the naked source are equivalent. However, for the naked  source the mapping is not exact and is only valid in the weak field approximation, as mentioned above. For the naked source the position of the ion maps the position of the Dirac particle instead of the redshift for the black hole, see \eqref{SM:Map1} versus Eq.~(5) in the main body of the paper.  

To sumarize, following a similar approach to that in references \cite{Lamata07,Bermudez07,Gerritsma10,Pedernales15} for the single-photon and \cite{Felicetti15} for the two-photon QRM, we can design a proposal for the quantum simulation of the DCS including a semiclassical gravity theory in the weak field limit using trapped ions as a quantum platform. Such a quantum simulation could be useful to study the free fall of a Dirac particle and to investigate the equivalence principle in a weak gravity field.

\section{Time evolution of expectation values of the position}\label{SM:TimeEvol}

In the trapped-ion implementation, the gravitational redshift maps to the position of the ion $\hat{X}=\frac{\lambda}{\sqrt{2}}(\hat{a}^{\dagger}+\hat{a})$. The position of the Dirac particle is related to the square of the position operator via $\hat{x} = (\hat{X}^2/r_s)+r_s$.  In this section, we are interested in calculating the time evolution of the redshift as well as of its squared value to make predictions on the trajectory of the Dirac particle as it approaches the event horizon, $x=\frac{1}{2M}=r_s$.  We first compute the Heisenberg equations for these operators and subsequently discuss the equations of motion for the corresponding expectation values. 

\subsection{Heisenberg equations}\label{SM:TimeEvol:Heisenberg}

First, let us compute the first order derivative of the operator $\hat{X}(t) = \hat{U}(t)^\dagger \hat{X} \hat{U}(t)$,
\begin{eqnarray}
\frac{d\hat{X}(t)}{dt}=\frac{i}{\hbar}[\hat{H}_D,\hat{X}(t)]&=&\frac{i \hat{U}(t)^\dagger [\hat{H}_D,\hat{X}]  \hat{U}(t) }{\hbar} \nonumber \\
&=&\frac{i}{\hbar} \hat{U}(t)^{\dagger}[\frac{c}{4r_s}  \{\hat{X},\hat{P}\}\hat{\sigma}_x + m c^2 \frac{\hat{X}}{r_s} \hat{\sigma}_z,\hat{X} ] \hat{U}(t) \nonumber \\
&=& \frac{i}{\hbar} \hat{U}(t)^{\dagger}[\frac{c}{4r_s} \{\hat{X},\hat{P}\}\hat{\sigma}_x,\hat{X}]  \hat{U}(t) 
\nonumber \\
&=& \frac{c}{2r_s}  \hat{U}(t)^{\dagger} (\hat{X} \hat{\sigma}_x)  \hat{U}(t)  \nonumber \\
&=& \frac{c}{2r_s}  \hat{X}(t) \hat{\Phi}_x(t) \ ,
\label{FirststepXt}
\end{eqnarray}
\noindent
where the scaling factor $\hat{\Phi}_x(t)= \hat{U}(t)^{\dagger}\sigma_x  \hat{U}(t)$ and $\hat{\Phi}_x(t)^2=\mathbb{I}$. We used $[\{\hat{X},\hat{P}\},\hat{X}^n]=-2^n i\hbar \hat{X}^n$ for $n\geq 1$ integer.   The general solution to this equation has the form
\begin{eqnarray}
\hat{X}(t)=\hat{X}(0) \exp\Big[\frac{c}{2 r_s} \int^{t}_{0} \hat{\Phi}_x(t')dt' \Big]\ .
\end{eqnarray}
\noindent
The time evolution of the scaling reads
\begin{eqnarray}
\frac{d\hat{\Phi}_x}{dt}&=&\frac{i}{\hbar}[\hat{H}_{D},\hat{\Phi}_x(t)] \nonumber \\
&=& \frac{i}{\hbar} \hat{U}(t)^{\dagger}[\hat{H}_{D},\hat{\sigma}_x] \hat{U}(t) \nonumber \\ 
&=& \frac{i}{\hbar}  \hat{U}(t)^{\dagger}[  m c^2 \frac{\hat{X}}{r_s} \hat{\sigma}_z,\hat{\sigma}_x] \nonumber \\
&=& \frac{i}{\hbar} \hat{U}(t)^{\dagger} mc^2 \frac{\hat{X}}{r_s}  \hat{U}(t)  \hat{U}(t)^{\dagger}\Big(+2i\hat{\sigma}_y \Big) \hat{U}(t) \nonumber \\
&=& -\frac{2mc^2}{\hbar r_s} \hat{X}(t)\hat{\Phi}_y(t)\ , \label{velocity}
\end{eqnarray}
\noindent
where $\hat{\Phi}_y(t)\equiv \hat{U}(t)^{\dagger}\hat{\sigma}_y \hat{U}(t)$.
The acceleration of $X$ is then given by
\begin{eqnarray}
\frac{d^2\hat{X}(t)}{dt^2}&=&\frac{c}{2r_s}\Big(\frac{d\hat{X}(t)}{dt}\hat{\Phi}_x(t)+\hat{X}(t)\frac{d\hat{\Phi}_x(t)}{dt}\Big) \nonumber \\
&=& \frac{c}{2r_s}\Big[\Big(\frac{c}{2r_s}  \hat{X}(t) \hat{\Phi}_x(t)\Big)\hat{\Phi}(t) - \hat{X}(t)^2  \frac{2mc^2}{\hbar r_s}\hat{\Phi}_y(t) \Big] \nonumber \\
&=& \frac{c^2}{4 r_s^2} \hat{X}(t) - \frac{mc^3}{\hbar r_s^2}\hat{X}(t)^2 \hat{\Phi}_y(t)\ .  \label{acceleration}
\end{eqnarray} 
\noindent
This is a nonlinear equation which likely cannot be solved analytically.  However, the interesting feature concerns the oscillations in the $X(t)^2$ term, indicating the possibility of Zitterbewegung of the gravitational redshift.  

Knowledge of the time evolution of $X$ provides information on the time evolution of the position of the particle, $ \frac{X^2}{r_s}=x-r_s$.  Using equations \eqref{velocity} and \eqref{acceleration}, we can find the velocity and acceleration of the simulated Dirac particle,
\begin{eqnarray}
\frac{d\left[\hat{X}(t)^2\right]}{dt}&=&\frac{i}{\hbar}[\hat{H}_{D},\hat{X}(t)^2]=\frac{i \hat{U}(t)^{\dagger}}{\hbar}[\hat{H}_{D},\hat{X}^2] \hat{U}(t) 
= \frac{c}{r_s}  \hat{U}(t)^{\dagger}\hat{X}^2 \hat{\sigma}_x  \hat{U}(t) =\frac{c}{r_s}\hat{X}(t)^2 \hat{\Phi}_x(t)\ , \label{xdot} 
\end{eqnarray}
leading to the Heisenberg equation of the operator $\hat{x}(t) \equiv \hat{U}^\dagger(t)\hat{x}\hat{U}(t) = \frac{\hat{X}(t)^2}{r_s}+r_s$  
\begin{equation}
\frac{d\hat{x}(t)}{dt}=\frac{1}{r_S}\frac{d\left[\hat{X}(t)^2\right]}{dt}  = \frac{c}{r^2_S} \hat{X}(t)^2 \hat{\Phi}_x(t) = c\Big(\frac{\hat{x}(t)}{r_S}-1\Big)\hat{\Phi}_x(t)\ .
\end{equation}

The second derivative of $\hat{x}(t)$ reads
\begin{equation}
\frac{d^2\hat{x}(t)}{dt^2}=\frac{1}{r_s}\frac{d^2 \hat{X}(t)^2}{dt^2}= \frac{c}{r_s^2}\left(\frac{d\left[\hat{X}(t)^2\right]}{dt} \hat{\Phi}_x(t) +\hat{X}(t)^2\frac{d\hat{\Phi}_x(t)}{dt}\right) = \frac{c}{r_s^2}\left(\frac{c}{r_s}\hat{X}(t)^2 \hat{\Phi}_x(t)^2 -\frac{2mc^2}{r_s\hbar}\hat{X}(t)^3 \hat{\Phi}_y(t)\right)  \ ,
\end{equation}
leading to
\begin{equation}
\frac{d^2\hat{x}(t)}{dt^2} = \frac{c^2}{r_s}\Big(\frac{\hat{x}(t)}{r_s}-1\Big)  - \frac{2mc^3 }{\hbar} \Big(\frac{\hat{x}(t)}{r_s}-1\Big)^{3/2}\hat{\Phi}_y(t)\ .
\end{equation}

To summarize, let us rewrite the previous Heisenberg equations
\begin{subequations}\label{SM:Heisenberg}
\begin{equation}\label{SM:Heisenberg1_X(t)}
\frac{d\hat{X}(t)}{dt} = \frac{c}{2r_s}  \hat{X}(t) \hat{\Phi}_x(t) \ ,
\end{equation}
\begin{equation}\label{SM:Heisenberg2_X(t)}
\frac{d^2\hat{X}(t)}{dt^2} =\frac{c^2}{4r_s^2}\hat{X}(t) - \frac{mc^3}{\hbar r_s^2}\hat{X}(t)^2 \hat{\Phi}_y(t)\ ,
\end{equation}
\begin{equation}\label{SM:Heisenberg1_x(t)}
\frac{d\hat{x}(t)}{dt} = c\Big(\frac{\hat{x}(t)}{r_S}-1\Big)\hat{\Phi}_x(t)\ ,
\end{equation}
\begin{equation}\label{SM:Heisenberg2_x(t)}
\frac{d^2\hat{x}(t)}{dt^2} = \frac{c^2}{r_s}\Big(\frac{\hat{x}(t)}{r_s}-1\Big)  - \frac{2mc^3 }{\hbar} \Big(\frac{\hat{x}(t)}{r_s}-1\Big)^{3/2} \hat{\Phi}_y(t)\ ,
\end{equation}
\end{subequations}
where $\hat{\Phi}_{x(y)}(t) \equiv \hat{U}^\dagger(t)\sigma_{x(y)}\hat{U}(t)$, with $\hat{\Phi}_{x(y)}(t)^2={\mathbb{I}}$.

\subsection{Expectation values}\label{SM:TimeEvol:ExpectValues}

To obtain the dynamical equations for the expectation values, it suffices to take the average with respect to the initial state $|\psi_i\ra=|\Psi_0\rangle $ of the equations \eqref{SM:Heisenberg}
\begin{subequations}\label{SM:Dynamics}
\begin{equation}\label{SM:FirstDer_X(t)}
\frac{d}{dt}\langle \hat{X}(t) \rangle= \frac{c}{2r_s}  \langle\hat{X}(t) \hat{\Phi}_x(t) \rangle\ ,
\end{equation}
\begin{equation}\label{SM:SecondDer_X(t)}
\frac{d^2}{dt^2} \langle \hat{X}(t) \rangle = \frac{c^2}{4r_s^2} \langle\hat{X}(t)\rangle -  \frac{mc^3}{\hbar r_s^2}\langle\hat{X}(t)^2 \hat{\Phi}_y(t)\rangle\ ,
\end{equation}
\begin{equation}\label{SM:FirstDer_x(t)}
\frac{d}{dt}\langle \hat{x}(t) \rangle = c\left\langle \Big(\frac{\hat{x}(t)}{r_S}-1\Big)\hat{\Phi}_x(t)\right\rangle \ ,
\end{equation}
\begin{equation}\label{SM:SecondDer_x(t)}
\frac{d^2}{dt^2}\langle \hat{x}(t) \rangle= \frac{c^2}{r_s}\Big(\frac{\langle \hat{x}(t)\rangle}{r_s}-1\Big)  - \frac{2mc^3 }{\hbar}\left\langle \Big(\frac{\hat{x}(t)}{r_s}-1\Big)^{3/2} \hat{\Phi}_y(t)\right\rangle\ .
\end{equation}
\end{subequations}
Equation~\eqref{SM:SecondDer_x(t)} shows the classical dynamical equation with an additional oscillatory term,  
\begin{equation}\label{SM:SecondDer_x(t)2}
\frac{d^2}{dt^2}\langle \hat{x}(t) \rangle= \frac{c^2}{r_s}\Big(\frac{\langle \hat{x}(t)\rangle}{r_s}-1\Big)  - \frac{2mc^3 }{\hbar}\left\langle \Big(\frac{\hat{x}(t)}{r_s}-1\Big)^{3/2} e^{2i\hat{H}_Dt/\hbar}\hat{\sigma}_y\right\rangle\ ,
\end{equation}
where we have used the anti-commutation relation between $\hat{\sigma}_y$ and $\hat{H}_D$ 
$$
\hat{H}_D\hat{\sigma}_{y} = \frac{c}{4r_s}\{\hat{X},\hat{P}\}\hat{\sigma}_x\hat{\sigma}_y+mc^2\frac{\hat{X}}{r_s}\hat{\sigma}_{z}\hat{\sigma}_{y} = -\hat{\sigma}_{y}\hat{H}_D\ .
$$

\subsection{Classical correspondence}\label{SM:TimeEvol:ClassCorr}

Let us recall the geodesic equation \cite{Mann91(2)} 
\begin{equation}\label{SM:Geo1}
\frac{d^2 x^{\gamma}}{d\lambda^2}+\Gamma^{\gamma}_{\mu\nu}\frac{dx^{\mu}}{d\lambda}\frac{dx^{\nu}}{d\lambda}\ ,
\end{equation}
where $\lambda$ is an affine parameter along the world line, $\Gamma^{\gamma}_{\mu\nu}$ are the Christoffel symbols, and 
\begin{equation}\label{SM:ds^2}
c^2d\tau^2 = g_{\mu\nu}dx^\mu dx^\nu = \alpha(x)c^2 dt^2 - \frac{dx^2}{\alpha(x)}\ ,
\end{equation} 
where $ds=cd\tau$ is the invariant spacetime interval and $\tau$ is the proper time.  
As mentioned previously we can alternatively use the polar coordinates $(t,X)$ which corresponds to the position of the ion. After rescaling the time and space variables as $t\mapsto t/t_s$ $X\mapsto X/r_s$, where $t_s = 2r_s/c$, one can write the spacetime interval as
\begin{equation}\label{SM:ds^2:Alt}
d\tau^2 = X^2 dt^2 - dX^2 \ ,
\end{equation}
where $\tilde{g}_{\mu\nu}(X) = \text{diag}(X^2,-1)$. The geodesic equation has the same general form as \eqref{SM:Geo1} after substituting $x$ by $X$. 

We note that the explicit geodesic equations \eqref{SM:Geo1} can also be derived from the variational principle 
\begin{equation}\label{SM:Action}
S = \int_{\tau_i}^{\tau_f} d\tau = \int_{t_i}^{t_f} dt \sqrt{X^2-\dot{X}^2} = \int_{\tau_i}^{\tau_f} d\tau \sqrt{X^2 (d_\tau t)^2 -(d_\tau X)^2} \ ,
\end{equation}
where $\dot{f}$ stands for $df/dt$ and $d_\tau f$ for $df/d\tau$. \\

\textit{Massless particle}. For the massless case,  $d\tau = 0$ and from Eq. \eqref{SM:ds^2} we find $\dot{x}^2 = c^2\alpha^2$ and 
\begin{equation}
\ddot{x}-\frac{1}{2}c^2\alpha\alpha'  = 0 \Rightarrow \ddot{x} = \frac{c^2}{2r_s^2} \left(x-r_s\right)\ ,
\end{equation}
leading to
\begin{equation}
x(t) = x_0 \pm r_s\left(e^{\pm t/\tau_s}-1\right)\ ,
\end{equation}
where $\tau_s \equiv c/r_s$ is a characteristic time of the Schwarzschild black hole and where the $\pm$ solutions depends on the initial velocity $\dot{x}(0) = \pm c$ pointing in or out of the black hole. 

\begin{figure}[]
\begin{center}
\includegraphics[width=0.35\columnwidth]{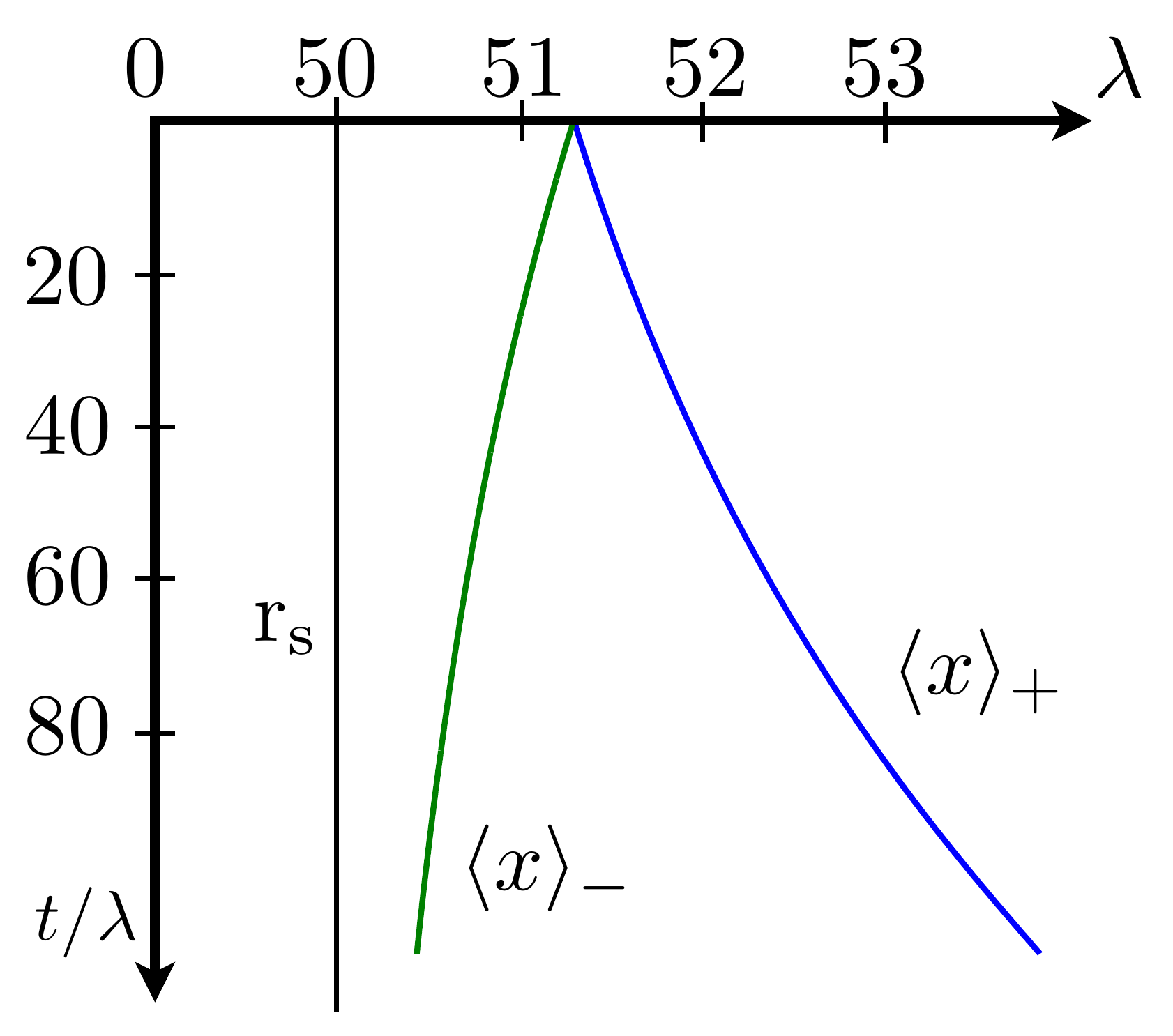}
\caption{{\bf Dynamics of the massless case.}  Plotted lines correspond to the time evolution of the expectation value of the position operator of a massless Dirac particle initially prepared in the internal state  $| + \rangle_x$ (right) and $| - \rangle_x$ (left). Here, $| \pm \rangle_x$ are the eigenstates of operator $\sigma_x$, and the spatial part of the initial wave wave function of the ion is $\phi_{X_0/\lambda=8}$, where $\phi_{X_0}$ is a Gaussian wave function centered at $X_0$. The simulation is performed under Hamiltonian in Eq.~(6) in the main text, for the case where $m=0$ and $M\lambda=0.01$. The vertical line labeled by $r_s$ indicates the position of the horizon. The chirality of the initial state defines whether the particle falls towards the horizon or escapes away from the origin.}
\label{massless}
\end{center}
\end{figure}

From Eq. \eqref{SM:FirstDer_x(t)} we can find these two scenarios. Assume the initial chirality is given by the eigenstate $|\pm\rangle_x$ of the Pauli matrix $\hat{\sigma}_x$ with eigenvalue $\pm 1$. Then, we find that the first order derivative of the expectation value of the position is given by
\begin{equation}\label{SM:MasslessEq}
\frac{d}{dt}\langle \hat{x}(t) \rangle = \pm c \left(\frac{\langle \hat{x}(t)\rangle}{r_s}-1\right)\ ,
\end{equation}
as $[\hat{H}_D,\hat{\sigma}_x]=0$ for the massless case, and $\langle \hat{\sigma}_x\rangle = \pm 1$. It follows that 
\begin{equation}\label{SM:MasslessMetric}
\left(\frac{d}{dt}\langle \hat{x}(t)\rangle \right)^2 = c^2\left(\frac{\langle \hat{x}(t)\rangle}{r_s}-1\right)^2 = c^2g_{00}\left[\langle \hat{x}(t)\rangle\right]\ ,
\end{equation}
which gives \eqref{SM:ds^2} with $g_{00}(x)=\alpha(x)$. In Fig. \ref{massless},  the numerical plot of the expectation value of the position of the ions is shown for positive and negative chirality, where the Hamiltonian is given in Eq.~(6) in the Letter. The trajectories clearly match the analytical solutions of Eq. \eqref{SM:MasslessEq}.  \\

\textit{Massive particle}. For the massive case, the geodesic equation is more complicated and it requires numerical analysis, see Fig.~(1) and the discussion in the main body of the paper. However, we can analyze qualitatively Eqs.~\eqref{SM:Dynamics}. For the sake of simplicity, we choose the variable $(t,X)$ in the following discussion and recall that the operator $c \hat{\Phi}_x(t) = c \hat{U}^\dagger(t)\sigma_x\hat{U}(t)$ is the velocity operator in Minkowski space \cite{McVittie1932}. Hence, in Eq. \eqref{SM:Heisenberg1_X(t)} the right hand side can be interpreted as the flat-space velocity multiplied by the redshift. After taking the expectation value, Eq. \eqref{SM:FirstDer_X(t)} determines the velocity of the particle in the reference frame of the stationary observer. Nevertheless, the term on the right hand side of this equation computes the expectation value of the operator $\hat{X}(t)\hat{\Phi}_x(t)$. To find the classical correspondence of this equation, we  split the expectation value of the product of these two operators by introducing the identity $\mathbb{I}=|\Psi_0\rangle\langle\Psi_0|+\hat{Q}\equiv \hat{P}_0+\hat{Q}$, where $|\Psi_0\rangle$ is the initial state. Therefore we can rewrite Eq. \eqref{SM:FirstDer_X(t)} as
\begin{equation}\label{SM:FirstDer_X(t)Class}
\frac{d}{dt}\langle \hat{X}(t) \rangle = \frac{c}{2r_s}\langle \hat{X}(t) \rangle \langle \hat{\Phi}_x(t) \rangle + \frac{c}{2r_s}\langle \hat{X}(t)\hat{Q}\hat{\Phi}_x(t) \rangle  \ .
\end{equation}
The first term in the right hand side of the previous equation corresponds to the classical equation while the second term takes into account the quantum interference. Now, taking the second derivative leads to
\begin{equation}\label{SM:SecondDer_X(t)Class}
\frac{d^2}{dt^2}\langle \hat{X}(t) \rangle = \frac{c}{2r_s}\frac{d}{dt}\langle \hat{X}(t)\rangle \langle \hat{\Phi}_x(t) \rangle +\frac{c}{2r_s}\langle \hat{X}(t) \rangle \frac{d}{dt}\langle \hat{\Phi}_x(t) \rangle  + \frac{c}{2r_s}\frac{d}{dt}\langle \hat{X}(t)\hat{Q}\hat{\Phi}_x(t) \rangle\ .
\end{equation}
The first term can be computed using  Eq. \eqref{SM:FirstDer_X(t)Class} leading to
$$
\frac{c^2}{4r_s^2}\langle \hat{X}(t)\rangle \langle \hat{\Phi}_x(t) \rangle ^2\ .
$$
The quantity $\langle \hat{\Phi}_x(t) \rangle ^2$ represents the square of the velocity in the local Minkowski reference frame. Using the conservation of the spacetime element $ds^2$ (in the system of units introduced in Eq. \eqref{SM:ds^2:Alt}) we find
\begin{equation}\label{EnergyCons}
ds^2 = X^2dt^2 - dX^2 = dT^2 - dX^2 \Rightarrow  v_x^2 = 1-X^2\ ,   
\end{equation}
where $T$ is the local Minkowski time and $v_x\equiv dX/dT$ the velocity measured in the local Minkowski frame. From the equation above, we deduce that the first derivative w.r.t. time $t$ of the velocity $v_x$ is $\dot{v}_x = - X^2$ which reads $\dot{v}_x = -\frac{c^2}{4r_s^2}X^2$ in the S.I. units. Therefore, the classical correspondence and Eq. \eqref{SM:SecondDer_X(t)Class} lead to the geodesic equation
\begin{equation}\label{SM:SecondDer_X(t)Class2}
\frac{d^2}{dt^2}\langle \hat{X}(t) \rangle = \frac{c^2}{4r_s^2}\langle \hat{X}(t)\rangle\left(1- \frac{\langle \hat{X}(t) \rangle^2}{r_s^2}\right) - \frac{c^2}{4 r_s^4}\langle \hat{X}(t)\rangle^3 + \text{quantum corrections}\ ,
\end{equation}
where the corrections contain quantum interference and mass-dependent terms (characterizing the Zitterbewegung effect).  To justify the classical correspondence, we use arguments developed in \cite{Arminjon13} where the relativistic energy conservation relation (derived classically from Eq. \eqref{EnergyCons}) is obtained from a semiclassical approximation. In this section we have shown that mass-dependent terms appear explicitly in the differential equation satisfied by the expectation value of the position operator, see Eqs. \eqref{SM:SecondDer_X(t)} and \eqref{SM:SecondDer_x(t)}, and they have no classical correspondence in Eq. \eqref{SM:SecondDer_X(t)Class2}.

To derive the classical geodesic equation for the polar coordinates $(t,X)$ (in S.I. units) we use Eqs. \eqref{SM:ds^2:Alt} and \eqref{SM:Action}
\begin{equation}\label{SM:ClassGeoX}
\ddot{X} = \frac{c^2}{4r_s^2}X - \frac{c^2}{2r_s^2} X^3 \ ,
\end{equation}
where the right hand side corresponds to the first term in Eq. \eqref{SM:SecondDer_X(t)Class2}.  
Using the coordinates $(t,x)$, we can rewrite this equation as
\begin{equation}\label{SM:ClassGeoxx}
\ddot{x} = \frac{c^2}{r_s}\left(\frac{x}{r_s}-1\right)-\frac{3c^2}{2r_s^2}x^2 \ .
\end{equation}

\subsection{Miscellaneous remarks}

\begin{figure}[]
\begin{center}
\includegraphics[width=0.4\columnwidth]{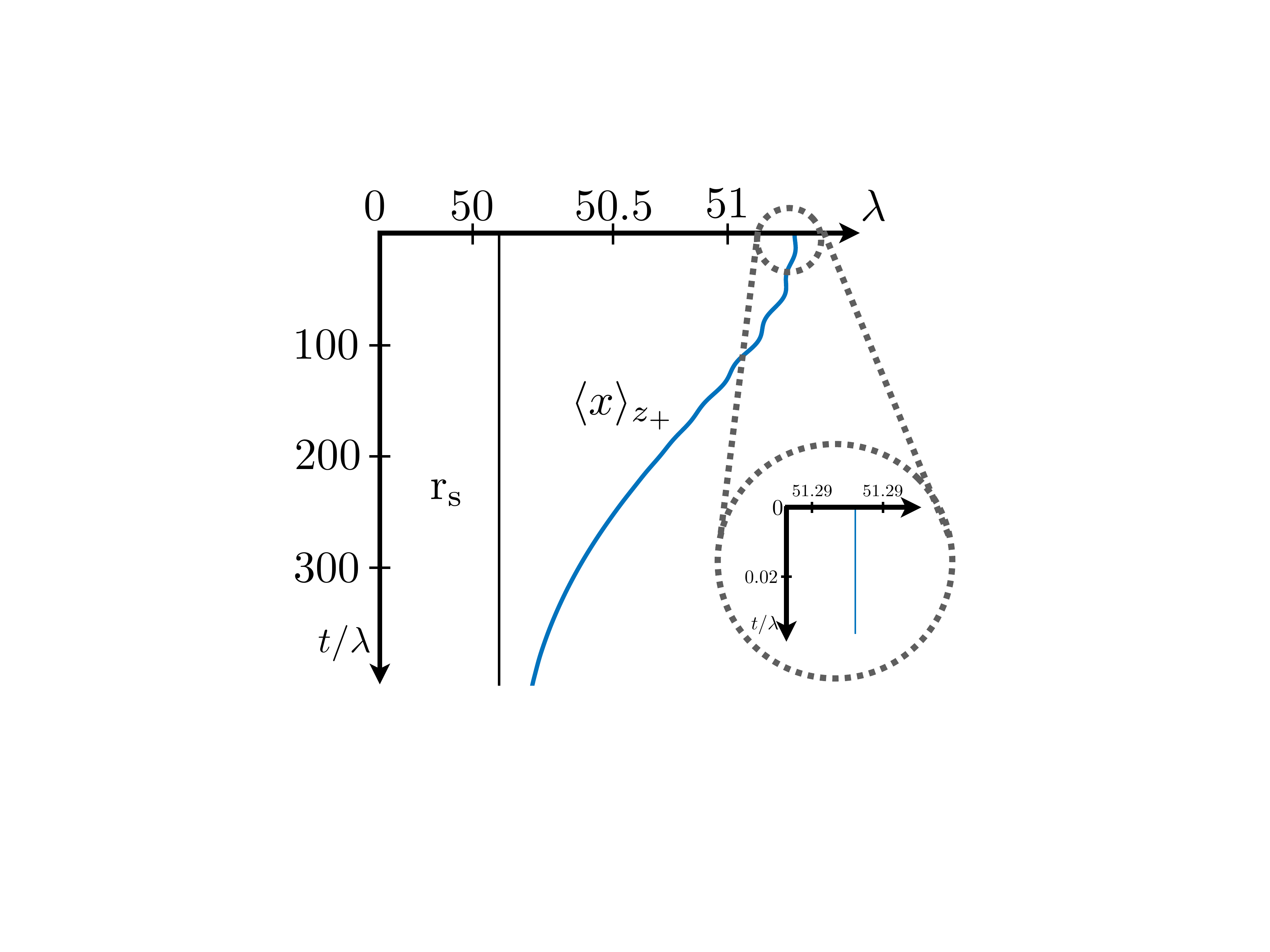}
\caption{{\bf Dynamics of the massive case for the initial state $|+\rangle_z$.} Time evolution of the expectation value of the position operator for an initial state $|+\rangle_z$ corresponding to an eigenvalue of the Pauli matrix $\sigma_z$. At short times (zoom), it is shown that the coefficient of the slope vanishes as predicted. Notice that the dynamics of the expectation value of the position operator for an initial state $|-\rangle_z$ is exactly the same. }
\label{massivesigmaz}
\end{center}
\end{figure}

Using our previous results, we can discuss a few complementary points:
\begin{itemize}
\item \textit{Slope at $t=0$.} We recall that the quantity $\langle\Psi_0| c\hat{\sigma}_x |\Psi_0\rangle$ can be interpreted as the velocity in a local Minkowski spacetime. Hence, for an initial internal state $ a|+\rangle_x + b|-\rangle_x$ we find the slope of the trajectories $\langle \hat{x}(t)\rangle$ and $\langle \hat{X}(t)\rangle$ at $t=0$ to be positive (negative) for $a>b$ ($a<b$) and zero for $a=b$, as the Minkowski velocity equals $c(a^2-b^2)$.  In Fig. \ref{massivesigmaz} we plot the trajectory $\langle \hat{X}(t)\rangle$ for $a=b$.   
\item \textit{Decay of Zitterbewegung effect.} When the particle approaches the horizon ( $X\rightarrow 0$), we observe numerically that the Zitterbewegung effect disappears. To prove this rigorously, we use Cauchy-Schwarz inequality in the second term of Eq. \eqref{SM:SecondDer_X(t)}
$$
\left|\langle \hat{X}(t)^2\hat{\Phi}_y(t) \rangle\right| 
\leq \sqrt{\langle \hat{X}(t)^4 \rangle} \sqrt{ \hat{\Phi}_y(t)^2  \rangle}
= \sqrt{\langle \hat{X}(t)^4 \rangle} \rightarrow 0\ , \ \text{when}\ t\rightarrow +\infty\ , 
$$ 
where we use that $\hat{\Phi}^\dagger_y(t)=\hat{\Phi}_y(t)$ and $\hat{\Phi}_y(t)^2 = 1$. 

\end{itemize}



\end{document}